\documentclass[aps,manuscript,superscriptaddress,]{revtex4}
\usepackage{graphicx}
\usepackage{amsmath}
\usepackage{amssymb}
\usepackage{colordvi}
\usepackage{mathrsfs}
\usepackage{bm}
\usepackage{verbatim}
\usepackage{dcolumn}
\usepackage{bm}
\usepackage{epsfig}
\usepackage{subfigure}

\begin{document}

\title{High-Efficiency Cooper-Pair Splitter in Quantum Anomalous Hall Insulator Proximity-Coupled with Superconductor}
\author{Ying-Tao Zhang}
\affiliation{College of Physics, Hebei Normal University, Shijiazhuang 050024, China}
\author{Xinzhou Deng}
\affiliation{International Center for Quantum Design of Functional Materials, Hefei National Laboratory for Physical Sciences at Microscale, and Department of Physics, University of Science and Technology of China, Hefei, Anhui 230026, China}
\author{Qing-Feng Sun}\thanks{Correspondence author: sunqf@pku.edu.cn}
\affiliation{International Center for Quantum Materials, School of Physics, Peking University, Beijing 100871, China}
\affiliation{Collaborative Innovation Center of Quantum Matter, Beijing 100871, China}
\author{Zhenhua Qiao}\thanks{Correspondence author: qiao@ustc.edu.cn}
\affiliation{International Center for Quantum Design of Functional Materials, Hefei National Laboratory for Physical Sciences at Microscale, and Department of Physics, University of Science and Technology of China, Hefei, Anhui 230026, China}

\date{\today}

\begin{abstract}
  \textbf{
  The quantum entanglement between two qubits is crucial for applications in the quantum communication. After the entanglement of photons was experimentally realized, much effort has been taken to exploit the entangled electrons in solid-state systems. Here, we propose a Cooper-pair splitter, which can generate spatially-separated but entangled electrons, in a quantum anomalous Hall insulator proximity-coupled with a superconductor. After coupling with a superconductor, the chiral edge states of the quantum anomalous Hall insulator can still survive, making the backscattering impossible. Thus, the local Andreev reflection becomes vanishing, while the crossed Andreev reflection becomes dominant in the scattering process. This indicates that our device can serve as an extremely high-efficiency Cooper-pair splitter. Furthermore, because of the chiral characteristic, our Cooper-pair splitter is robust against disorders and can work in a wide range of  system parameters. Particularly, it can still function even if the system length exceeds the superconducting coherence length.
}
\end{abstract}

\maketitle

The crossed Andreev reflection~\cite{Byers, Deutscher, Falci, Recher}, also known as non-local Andreev reflection, describes the process of converting an electron incoming at one terminal into an outgoing hole at another spatially-separated terminal.
By making use of the crossed Andreev reflection, a Cooper-pair in the superconductor can be split into two electrons, which propagate at two spatially-separated terminals while keeping their spin and momentum entangled.
These spatially-separated entangled electrons are the key building blocks for solid-state Bell-inequality experiments, quantum teleportation and quantum computation~\cite{Beckmann, Russo, Nielsen, Lesovik, Chtchelkatchev,
Bayandin}.
Therefore, the crossed Andreev reflection has received intensive attention in the past decade, and some crossed Andreev reflection-based Cooper-pair splitters have been proposed, \textit{e.g.}, a superconductor junction coupled with a quantum dot~\cite{Recher, Recher1, Hofstetter, Hofstetter1}, Luttinger liquid wires~\cite{Recher2}, carbon nanotubes~\cite{Bena, Herrmann}, and graphene~\cite{Cayssol, Wang}.
Recently, due to the quick emergence of the 2D $\mathbb{Z}_2$ topological insulators (TIs) accompanying with odd pairs of spin-helical counter-propagating edge modes along each boundary~\cite{Kane, Bernevig, Wiedmann,Roth,qiao-TI,DuRR}, some Cooper-pair splitters based on the TIs have been proposed.
For example, a TI-based Cooper-pair splitter was used to test the Bell inequality on solid state spins~\cite{Chen} and an all-electric TI-based Cooper-pair splitter was proposed with crossed Andreev reflected hole being spatially separated from the tunneled electron~\cite{Reinthaler}.
On the experimental side, the crossed Andreev reflection and the Cooper-pair splitting have been confirmed in quantum dot systems~\cite{Hofstetter,Hofstetter1}, carbon nanotubes~\cite{Herrmann}, \textit{etc}.

However, so far all the reported Cooper-pair splitters inherently exhibit various disadvantages.
First, since the incoming electrons and outgoing holes in the crossed Andreev reflection reside in spatially separated terminals, the coefficient of the crossed Andreev reflection is usually rather limited and decays exponentially with the increase of distance between the two terminals.
Second, some proposed Cooper-pair splitters can only work under certain special system parameters and are usually not robust against the disorders.
Thus the crossed Andreev reflection is strongly decreased in the presence of disorders and impurities.
Third, the local Andreev reflection often occurs inevitably and dominates the scattering process, \textit{e.g.}, for the representing TI-based splitters, the helical edge states give rise to a sizeable Andreev reflection in such hybrid systems~\cite{Reinthaler, Sun, Narayan}, which leads to the weak crossed Andreev reflection and the very low-efficiency Cooper-pair splitting.

Inspired by the exotic chirally propagating transport properties of the quantum anomalous Hall insulator (QAHI), we propose a Cooper-pair splitter in
the hybrid system by coupling the QAHI with a superconductor. Particularly, the proposed Cooper-pair splitter can overcome all the above mentioned weaknesses.
QAHI is a special realization of the quantum Hall effect~\cite{QHE1,QHE2} that occurs in the absence of an external magnetic field, in which the chiral edge states protected by the spatial-separation allow the dissipationless current transport in 2D electronic systems. This effect has been theoretically proposed in various systems~\cite{proposal0,proposal1,Liu,proposal3,Yu,Qiao,proposal6,proposal7,proposal8,proposal9,proposal10,proposal11,proposal12}, but was first realized in TI thin films by introducing the intrinsic ferromagnetism to break the time-reversal symmetry~\cite{Qi1, Qi2, Liu, Yu, Nomura,ChangCuiZu,ExperimentalQAHE1,ExperimentalQAHE2}.
Therefore, given the absence of backscattering of the quantum anomalous Hall edge modes, it is reasonable to expect that the local Andreev reflection is forbidden and the crossed Andreev reflection could be considerably improved in the hybrid structure composed of the QAHI and superconductor.

\bigskip
\textbf{\large Results}
\bigskip

\begin{figure*}[t]
  \includegraphics[width=15cm]{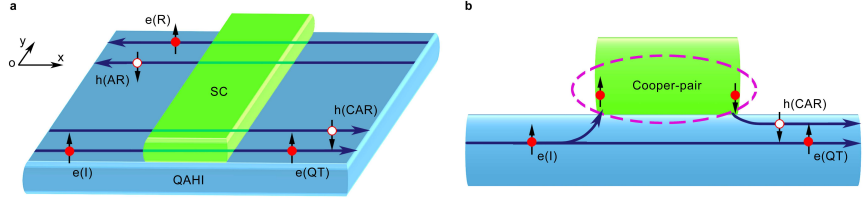}
  \caption{\textbf{Schematic of the proposed Cooper-pair splitter.}
  \textbf{a.} Schematic of a QAHI covered with a superconductor (labelled as ``SC") in the central region. When an electron with up-spin [marked as ``e(I)"] incoming from the left terminal enters the central hybrid region, there correspond four different scattering processes: (1) Direct reflection as a spin-up electron to the left terminal [marked as ``e(R)"]; (2) Local Andreev reflection as a spin-down hole to the left terminal [marked as ``h(AR)"]; (3) Quantum tunnelling as a spin-up electron to the right terminal [marked as ``e(QT)"]; and (4) Crossed Andreev reflection as a spin-down hole to the right terminal [marked as ``h(CAR)"].
  \textbf{b.} Side-view of the schematic displayed in \textbf{a.} It illustrates that the incoming electron propagates through the central region in two ways: (1) directly tunnelling through the QAHI as an electron, and (2) proximity-flowing into the SC to form a spatially separated Cooper-pair and ejecting a hole to the right terminal. Red solid and empty circles are used to denote the electrons and holes, respectively.} \label{fig1}
\end{figure*}

\textbf{System Model}.--- In this article, we study the quantum tunnelling, Andreev reflection and crossed Andreev reflection in a two-terminal finite-sized QAHI system, with the central region being covered by a superconductor (See Fig.~\ref{fig1}).  The total Hamiltonian of the hybrid system can be written as $H_{\rm T}=H_{\rm QAHI}+H_{\rm SC}+H_{\rm C}$, where $H_{\rm QAHI}$, $H_{\rm SC}$, and $H_{\rm C}$ correspond respectively to the Hamiltonians of the QAHI, superconductor, and their coupling. As a concrete example, the QAHI is modelled by using a monolayer graphene including the Rashba spin-orbit coupling and an exchange field, and its tight-binding Hamiltonian can be written as~\cite{Qiao}:
\begin{eqnarray}
H_{\rm QAHI}=-t\sum_{<ij>\alpha}c^{\dagger}_{i\alpha}c_{j\alpha}+\lambda\sum_{i\alpha}c^{\dagger}_{i\alpha}\sigma_{z}c_{i\alpha}
+ t_{\rm R}\sum_{<ij>\alpha\beta} \textit{\textbf{e}}_{z}\cdot({\sigma}_{\alpha\beta}\times \bm{d}_{ij})c^{\dagger}_{i\alpha}c_{j\beta}+\sum_{i\alpha}\epsilon_{i}c^{\dagger}_{i\alpha}c_{i\alpha},
\end{eqnarray}
where $t$ measures the nearest-neighbor hopping amplitude that is set as the unit of energy, and $c^{\dagger}_{i\alpha}$ ($c_{i\alpha}$) is the electron creation (annihilation) operator at site $i$ with spin $\alpha$ (\textit{i.e.,} $\uparrow$ or $\downarrow$). The second term corresponds to the exchange field with a strength of $\lambda$, and $\mathbf{\sigma}$ are the spin-Pauli matrices. The third term describes the external Rashba spin-orbit coupling with a coupling strength of $t_{\rm R}$, arising from the mirror symmetry breaking, \textit{e.g.,} by applying a vertical electric field~\cite{Kane}. Here, $\bm{d}_{ij}$ is a unit vector pointing from site $j$ to site $i$. In the last term, the static Anderson-type disorder is added to $\epsilon_{i}$ with a uniform distribution in the interval of [-$W$/2, $W$/2], where $W$ characterizes the strength of the disorder. In addition, the Hamiltonians of the superconductor and its coupling with the QAHI can be respectively expressed as:
\begin{eqnarray}
H_{\rm SC}&=&\sum_{\bm{k}, \alpha}\epsilon_{\bm{k}} b^{\dagger}_{\bm{k} \alpha}b_{\bm{k} \alpha}+
\sum_{\bm{k}}\Delta(
b^{\dagger}_{\bm{k} \uparrow}b^{\dagger}_{-\bm{k}\downarrow}+
b_{-\bm{k} \downarrow}b_{\bm{k} \uparrow}),\\
H_{\rm C} &= &-t_{\rm C}\sum_{i,\alpha}c^{\dagger}_{i\alpha}b_{\alpha}(\bm{r}_i)+h.c.
\end{eqnarray}
where $\epsilon_{\bm{k}}$ corresponds to the on-site energy in the momentum space, $\bm{k}=(k_x,k_y)$ is the wave vector,
$\Delta_{\rm S}$ is the superconducting pair-potential measuring the superconductor gap, $t_{\rm C}$ is the hopping amplitude between the superconductor and the QAHI, and $b_{\alpha}(\bm{r})=\sum_{\bm{k}}e^{i\bm{k}\cdot{\bm{r}}}b_{\bm{k}\alpha}$ is the annihilation operator at the position $\bm{r}$ in the real space. The size of the central region is denoted by $N\times L$, where $L$ and $N$ count the atom numbers along $x$ and $y$ directions, respectively.

\begin{figure*}[t]
  \includegraphics[width=12cm]{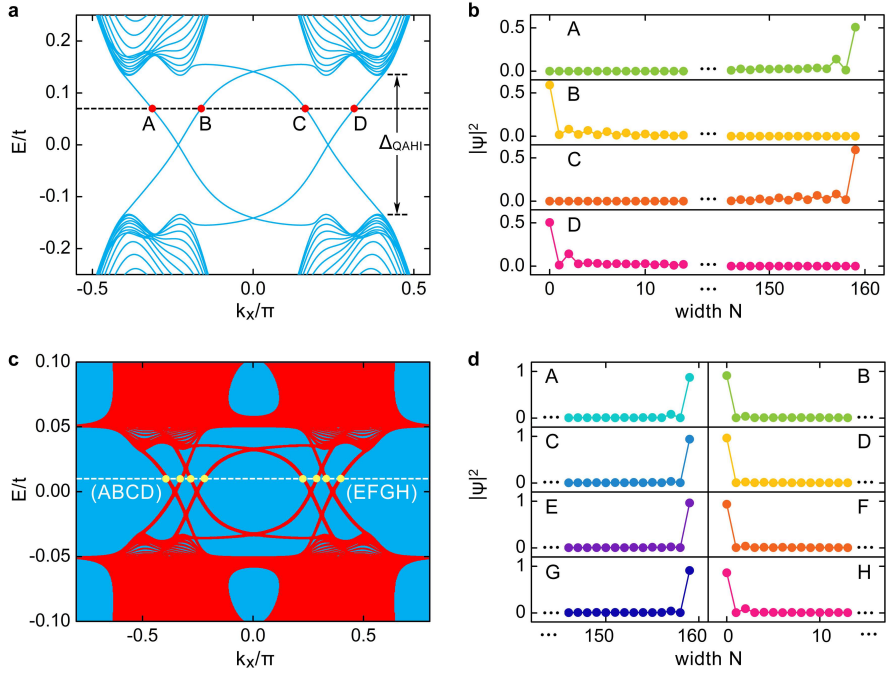}
  \caption{
  \textbf{The band structures of the QAHI un-covered and covered by the superconductor.}
  \textbf{a.} Band structure of a zigzag graphene nanoribbon with Rashba spin-orbit coupling $t_{\rm R}=0.20 t$ and Zeeman field $\lambda=~0.18 t$. There are four different states labelled as ``A"- ``D" for any fixed Fermi level inside the bulk gap. \textbf{b.} Wave-function distributions $|\psi|^{2}$ across the width for the four states labelled in \textbf{a}. Only part of the ribbons are shown.
  States ``A" and ``C"  are localized at the top boundary, while states ``B" and ``D" are localized at the low boundary.
  \textbf{c.} The spectral function $A(E, k_{x})$ for the hybrid system of QAHI coupled with a grounded superconductor. The parameters of the superconductor are set to be $\Delta=0.05t$ and $g_{s}=2\Delta$. There are eight different edge states ``A"-``H" for the Fermi level inside the superconductor gap. \textbf{d.} Wave-function distributions $|\psi|^{2}$ across the width for the eight states labelled in \textbf{c}. States ``A", ``C", ``E", and ``G"  are localized at the top boundary, while states ``B", ``D", ``F", and ``H" are localized at the low boundary.
  } \label{fig2}
\end{figure*}

\bigskip
{\textbf{Physical picture of  the absence of the normal and Andreev reflections}---.} In the absence of the superconductor, gapless edge modes of the QAHI appear inside the bulk band gap $\Delta_{\rm QAHI}$ of the graphene nanoribbon (See Fig.~\ref{fig2}\textbf{a}). For a given Fermi-level lying inside the bulk band gap, \textit{e.g.}, the crossing points by the dashed line displayed in Fig.~\ref{fig2}\textbf{a}, there correspond four different electron states labelled as ``A", ``B", ``C", and ``D". In Fig.~\ref{fig2}\textbf{b}, we plot the wavefunction distributions of these states across the width. It can be clearly seen that the wavefunction of each state is mainly localized at the ribbon boundary, \textit{i.e.}, states ``A" and ``C" are localized at the top boundary, while states ``B" and ``D" are localized at the bottom boundary, which confirms that they are indeed the edge states. From the dispersion relation shown in Fig.~\ref{fig2}\textbf{a}, one can also find that states ``A" and ``C" propagate along the same direction, which is opposite from that of states ``B" and ``D". These together indicate the unidirectionally or chirally propagating property of the edge states, which is distinct from the helical edge states of the $\mathbb{Z}_2$ topological insulators. For clarity, the chiral edge states are visually displayed in Fig.~\ref{fig1}\textbf{a} with the blue arrows signifying the propagating directions.

We now study how the chiral edge states are affected when the QAHI is covered by a superconductor. Usually, when a conductor or the helical edge states of TIs are covered by a superconductor, a gap can open at the Fermi surface due to the proximity effect from the coupling with the superconductor. The reason behind is that, for the normal conductors and the $\mathbb{Z}_2$ TIs, the dispersion relation of opposite spin states is usually an even function of momentum $\bm{k}$ due to the time-reversal symmetry, \textit{e.g.} $\epsilon_{\bm{k}\uparrow} = \epsilon_{-\bm{k}\downarrow}$.
In such cases, a superconducting pair-potential $\Delta_{\rm S}$ can open a band gap and the energy spectra become $E_{\bm{k}} = \pm \sqrt{\epsilon_{\bm{k}\uparrow}^2 +\Delta_{\rm S}} - E_{\rm F}$.
However, it is rather different for the situation of the chiral edge states of the QAHI.
Fig.~\ref{fig2}\textbf{c} shows the spectral function $A(E, k_{x})$ of the hybrid system of QAHI covered by a superconductor, where
$A(E, k_{x})=-1/{\pi}{\rm{Im}}\{{\rm{Tr}}[{\bf G}^{r}(E, k_{x})]\}$ with ${\bf G}^{r}(E, k_{x})$ being the retarded Green's function [See METHODS for the calculation details].
One can see that at any fixed energy $E$, the spectral function has a finite value (denoted in red), which reflects that no band gap opens at the Fermi surface in the hybrid system.
The underlying reason can be attributed to that the dispersion relation $\epsilon^{\rm edge}_{k_x\sigma}$ of the chiral edge modes with opposite spins is an odd function of the momentum $k_x$ in the QAHI, \textit{i.e.}, $\epsilon^{\rm edge}_{k_x\uparrow} = -\epsilon^{\rm edge}_{-k_x\downarrow}$, as shown in Fig.~\ref{fig2}\textbf{a}. In this case, if a superconducting pair potential $\Delta_{\rm S}$ is applied to the chiral edge states with a Hamiltonian of $\tilde{H}=\sum_{k_x, \sigma} (\epsilon^{\rm edge}_{k_x\sigma} -E_{\rm F})b^{{\rm edge}, \dagger}_{k_x\sigma} b^{\rm edge}_{k_x\sigma} +\sum_{k_x} \tilde{\Delta}[ b^{{\rm edge}, \dagger}_{k_x\uparrow} b^{{\rm edge},\dagger}_{-k_x\downarrow} + b^{{\rm edge}}_{-k_x\downarrow} b^{{\rm edge}}_{k_x\uparrow}]$, the energy bands become $E_{k_x} =\epsilon_{k_x \uparrow} -E_{\rm F} \pm \Delta_{\rm S}$. This new energy dispersion indicates that the chiral edge states still keep gapless, and thus no band gap opens. By analyzing the wavefunction distributions $|\psi|^{2}$ across the width of the QAHI ribbon covered with a superconductor as displayed in Fig.~\ref{fig2}\textbf{d},
one can find that all electron states inside the superconductor gap are localized at the system boundaries.
For example, the states ``A", ``C", ``E", and ``G" propagating from right to left are localized at the top boundary whereas the states ``B", ``D", ``F", and ``H" propagating from left to right are localized at the low boundary, which means that the edge states exhibit the chiral propagating characteristic.
Therefore, the backscattering is completely forbidden, because that the chiral edge modes are topologically protected by the spatial separation.

We now turn to analyze the scattering processes when an electron with fixed spin (\textit{e.g.}, up spin) incoming from the left terminal flows into the central region of the hybrid system. In general, there are four scattering processes as displayed in Fig.~\ref{fig1}\textbf{a}: 1) the direct reflection to the left terminal as a spin-up electron; 2) the quantum tunnelling to the right terminal as a spin-up electron; 3) the Andreev reflection to the left terminal as a spin-down hole; and 4) the crossed Andreev reflection to the right terminal as a spin-down hole. To be specific, in Fig.~\ref{fig1}\textbf{b}, we present a schematic illustration of how the Andreev reflection and the crossed Andreev reflection occur in the central hybrid region.
Note that the incoming electron from the left terminal propagates along the bottom boundary, while for the direct reflection and the Andreev reflection, the outgoing electron and hole propagate along the top boundary (see Fig.~\ref{fig1}\textbf{a}). Furthermore, there exists a bulk band gap no matter whether the QAHI is covered or not by a superconductor, thus the scattering between the top and bottom boundaries are almost impossible for a wide enough ribbon. Therefore it is reasonable to expect that the direct reflection and Andreev reflection will be completely suppressed, while the quantum tunneling and the crossed Andreev reflection will dominate the whole scattering processes. In other words, the two electrons of a Cooper-pair go respectively to the left terminal and right terminal as described in Fig.~\ref{fig1}\textbf{b}, which leads to a high efficiency of the Cooper-pair splitting.

\begin{figure}[t]
  \includegraphics[width=8cm]{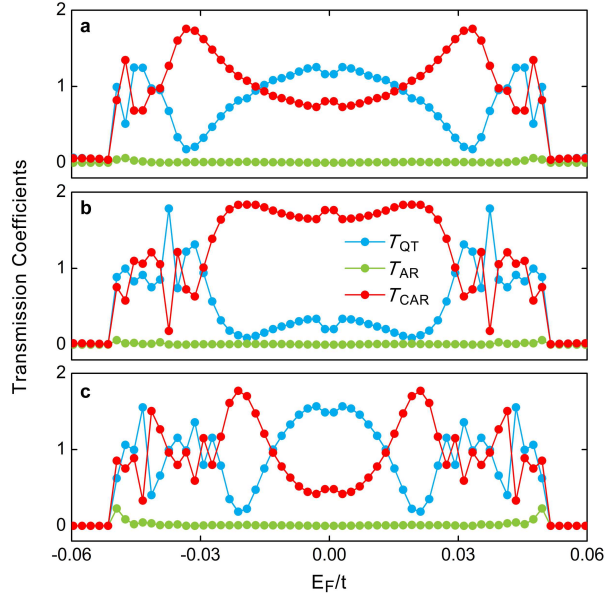}
  \caption{\textbf{Transmission coefficients as functions of the Fermi level.}
  Transmission coefficients of the quantum tunneling $T_{\rm QT}$,
  Andreev reflection $T_{\rm AR}$, and crossed Andreev reflection $T_{\rm CAR}$ as functions of the Fermi level $E_{\rm F}$ for different system lengths $L=11$ (\textbf{a}), 21 (\textbf{b}) and
  31 (\textbf{c}) at a fixed system width of $N=80$. Other parameters are the same as
  those in Fig.~\ref{fig2}\textbf{c}.}
  \label{fig3}
\end{figure}

\textbf{Numerical results and discussions}---. In this Section, we provide a detailed numerical calculation to support our above expectation. Hereinbelow, the parameters for the QAHI are chosen to be $\lambda=0.18~t$ and $t_{\rm R}=0.20~t$, and the pair potential of the superconductor is set to be $\Delta_{\rm S}=0.05~t$. It is noteworthy that the size of the QAHI bulk band gap $\Delta_{\rm QAHI} \approx 0.26~t$ is much larger than the superconducting gap $2\Delta_{\rm S}$.
Figure~\ref{fig3} plots the transmission coefficients of the quantum tunneling $T_{\rm QT}$, Andreev reflection $T_{\rm AR}$ and crossed Andreev reflection $T_{\rm CAR}$ as functions of the Fermi level $E_{\rm F}$, where the width of the ribbon is fixed at $N=80$ that is wide enough to avoid the finite-size effect and the lengths are respectively chosen to be $L=11$, 21, and 31 (see METHODS for the calculation details).
One can find that, as expected, in all the three different cases the Andreev reflection is completely suppressed to be $T_{\rm AR}\approx 0$ regardless of the length $L$, as long as the Fermi-level lies inside the superconducting gap $|E_{\rm F}|<\Delta_{\rm S}$.
Because of the absence of both the direct reflection and the Andreev reflection, the electrons incoming from the left terminal propagate into the right terminal in forms of the quantum tunneling and the crossed Andreev reflection, leading to $T_{\rm QT} +T_{\rm CAR}\approx 2$ as long as $|E_{\rm F}| <\Delta_{\rm S}$.
Most importantly, the vanishing of the Andreev reflection also results in an extremely high Cooper-pair splitting efficiency $\eta\rightarrow100\%$ independent of the system length $L$ as displayed in Fig.~\ref{fig3}, which is defined as $\eta=T_{\rm CAR}/(T_{\rm AR}+T_{\rm CAR})$.
This strongly suggests that our proposed system can function as a high-efficiency Cooper-pair splitter.
In addition, for the electrons incoming from two opposite spin edge states, they have the same crossed Andreev reflection. This means that the spin-up (spin-down) electron has the same probability to go to left or right terminal. And if the spin-up electron goes to the left terminal, then the spin-down one has to go to the right terminal. So the two spatially-separated electrons from a Cooper-pair still keep the spin and momentum entangled. Another observation in Fig.~\ref{fig3} is that the transmission coefficient of the quantum tunnelling $T_{\rm QT}$ is always comparable with that of the crossed Andreev reflection $T_{\rm CAR}$, exhibiting a universal characteristic for different system sizes whenever the Fermi energy satisfies $|E_{\rm F}<\Delta_{\rm S}|$.
To eliminate the influence of the quantum tunnelling for practical applications, one can simply apply the same potential in both the left and right terminals. Then, the crossed Andreev reflection can be utilized to design high-efficiency Cooper-pair splitters by mediating the potential of the central region.

\begin{figure}[tbp]
  \includegraphics[width=10cm]{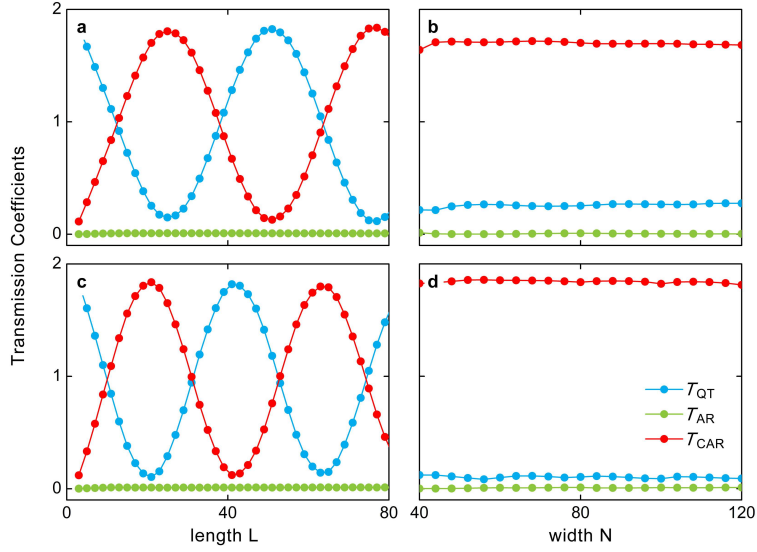}
  \caption{
  \textbf{The relations of transmission coefficients with the device size.}
  \textbf{a} and \textbf{c}: Transmission coefficients of the quantum tunneling $T_{\rm QT}$, Andreev reflection $T_{\rm AR}$, and crossed Andreev reflection $T_{\rm CAR}$ as functions of system length $L$ for different Fermi levels $E_{\rm F}=0.01t$ (\textbf{a}) and $0.02t$ (\textbf{c}) at the fixed width of $N=80$. \textbf{b} and \textbf{d}: The transmission coefficients as functions of the system width $N$ for different Fermi levels $E_{\rm F}=0.01t$ (\textbf{b}) and $0.02t$ (\textbf{d}) at the fixed system length of $L=21$. Other parameters are the same as those in Fig.~\ref{fig2}\textbf{c}.}
\label{fig4}
\end{figure}

Next, we move to the size dependence of these transmission coefficients at fixed Fermi-levels. Figures~\ref{fig4}\textbf{a} and \ref{fig4}\textbf{c} display the transmission coefficients of the quantum tunnelling, Andreev reflection, and crossed Andreev reflection as functions of the system length $L$ at a fixed system width $N$.
One can find that the Andreev reflection is also vanishing with $T_{\rm AR}=0$ for all values of $L$, while the transmission coefficients $T_{\rm QT}$ and $T_{\rm CAR}$ oscillate as functions of the system length $L$ with the oscillation period being dependent on the Fermi-level $E_{\rm F}$.
The reason behind this observation is that the covered superconductor functions as an applied external potential, and the resulting quantum tunnelling and crossed Andreev reflection can reach a resonance at certain system lengths.

It is noteworthy that the difficulty in realizing Cooper-pair splitter is that the crossed Andreev reflection is intimately affected by
the distance between the two normal terminals. For all previous proposed Cooper-pair splitter, the crossed Andreev reflection quickly decreases in parallel to increasing the system length and finally vanishes when it exceeds the superconducting coherence length. Counterintuitively, in our considered system the crossed Andreev reflection can still survive and keep a large value even for relatively long system lengths. And the obtained coefficient of the crossed Andreev reflection is larger than 0.1 for any system lengths $L$ (the distance between the two normal terminals). This is perfectly logical and reasonable, because the chiral edge states exist in the QAHI no matter whether it is covered or not by the superconductor, making the scattering from one boundary to other one almost impossible (except for very narrow ribbons). Therefore, our proposed high-efficiency Cooper-pair splitter is able to function in a long-range junction, even if it farther exceeds the superconducting coherence length. In addition, by choosing proper system lengths or externally adjusting Fermi-levels, our proposed setup can not only reach a high Cooper-pair splitting efficiency $\sim100\%$, but also provide a strong signal of the crossed Andreev reflection with $T_{\rm CAR}\approx 2$, which is much larger than those reported in previous works.

Figures~\ref{fig4}\textbf{b} and \ref{fig4}\textbf{d} show the transmission coefficients of the quantum tunneling, Andreev reflection and crossed Andreev reflection as functions of the sample width $N$ at fixed system length $L=21$.
One can see that only for small width $N$ there exists very weak Andreev reflection, because, in this case, both electrons and holes can be scattered between the top and bottom boundaries of the device.
In cases with larger $N$, the edge states at the two boundaries are well separated, leading to the disappearance of the Andreev reflection $T_{\rm AR}=0$ and the saturation of $T_{\rm QT}+T_{\rm CAR}\approx 2$.

\begin{figure}[t]
  \includegraphics[width=10cm]{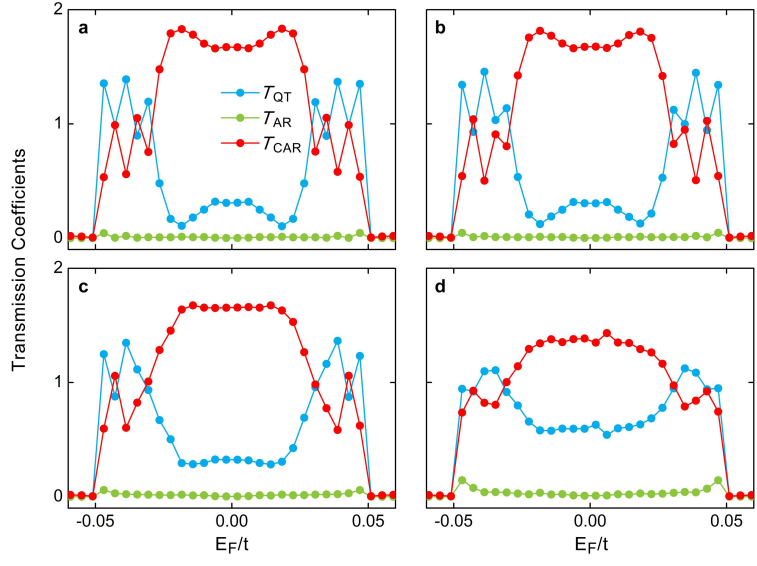}
  \caption{\textbf{The effect of disorders on transmission coefficients.}
  Averaged transmission coefficients for quantum tunneling $T_{\rm QT}$, Andreev reflection $T_{\rm AR}$, and crossed Andreev reflection $T_{\rm CAR}$ as functions of the Fermi level $E_{\rm F}$ for different disorder strengths $W/\Delta=1$ (\textbf{a}), 2 (\textbf{b}), 5 (\textbf{c}) and 10 (\textbf{d}). Other parameters are the same as those used in Fig.~\ref{fig2}\textbf{c}. Over 200 samples are collected for the average at each point.}
  \label{fig5}
\end{figure}

At this point, we have proposed a scheme for realizing a high-efficiency Cooper-pair splitter in a hybrid system of the QAHI proximity-coupled with a superconductor. It is clear that external disorders are inevitable in practical devices. Therefore, the question naturally arises as to whether the quantum tunnelling and the crossed Andreev reflection are still robust in real systems? To address this issue, in Fig.~\ref{fig5} we plot the averaged transmission coefficients by collecting over 200 samples in the presence of on-site Anderson disorders in the central scattering region. One can see that these transmission coefficients are nearly unaffected when the relatively strong disorders of $W/\Delta_{\rm S}=1$, 2 and $5$ are introduced. Even for a much stronger disorder strength of $W/\Delta_{\rm S}=10$, the Andreev reflection is only slightly enhanced. This is because the electrons can be weakly scattered to the opposite edges in the presence of rather strong disorders. Therefore, even in such a case, the quantum tunneling and the crossed Andreev reflection still dominate the whole scattering process. All these observations demonstrate that our proposed high-efficiency Cooper-pair splitter is much robust against external disorders, indicating its experimental feasibility. It should also be noted that although we have used a graphene-based QAHI as a specific example in this article, our findings can be applied to any other systems that can realize the quantum anomalous Hall effect.

{\bf\large Methods}

In our numerical calculation, we have mainly employed the non-equilibrium Green's function technique~\cite{Datta} and the recursive transfer matrices method to compute various transmission coefficients in a two-terminal mesoscopic system. For example, the transmission coefficients for the quantum tunneling, the Andreev reflection, and crossed Andreev reflection are expressed as~\cite{sun1,sun2}:
\begin{eqnarray}
{T}_{\rm QT}&=& {\rm{Tr}}[{\bf \Gamma}^{\rm L}_{ee}{\bf G}^{r}_{ee}{\bf\Gamma}^{\rm R}_{ee}{\bf G}^{a}_{ee}],  \\
{T}_{\rm AR}&=& {\rm{Tr}}[{\bf \Gamma}^{\rm L}_{ee}{\bf G}^{r}_{eh}{\bf\Gamma}^{\rm L}_{hh}{\bf G}^{a}_{he}],  \\
{T}_{\rm CAR}&=&{\rm{Tr}}[{\bf\Gamma}^{\rm L}_{ee}{\bf G}^{r}_{eh}{\bf\Gamma}^{\rm R}_{hh}{\bf G}^{a}_{he}],
\end{eqnarray}
where ``\textit{e}/\textit{h}" is an abbreviated expression of ``electron/hole", and ``L/R" indicates ``left/right". ${\bf G}^{r}(E)=(E{\bf I}-{\bf H}-{\bf\Sigma}^{r}_{\rm L}-{\bf\Sigma}^{r}_{\rm R}-{\bf\Sigma}^r_{\rm S})^{-1}$ is the retarded Green's function, where ${\bf H}$ stands for the Hamiltonian of the central region in the Nambu space, ${\bf\Sigma}^{r,a}_{\rm L/R}$ are the self-energies of the left/right terminals and can be numerically obtained~\cite{Lee},
and ${\bf\Sigma}^r_{S}={\bf t}_{\rm C}{\bf g}_{\rm S}^{r}{\bf t}_{\rm C}^{*}$ is the self-energy of the superconductor terminals with ${\bf g}_{\rm S}^{r}$ representing the surface Green's function of the semi-infinite superconductor terminal
that equals to the bulk Green's function for the conventional s-wave superconductor~\cite{Samanta}. In our calculations, we take the self-energy of the superconductor terminal $\Sigma^r_{{\rm S}, ij}=-{i\delta_{ij}g_{\rm s}}/(2\Omega)\left[
\begin{array}{cc}
1 & \Delta_{\rm S}/E \\
\Delta_{\rm S}/E & 1 \\
\end{array}
\right] $, where $\Omega=\sqrt{E^{2}-\Delta_{\rm S}^{2}}/|E|$ when
$|E|>\Delta_{\rm S}$ and $\Omega=i\sqrt{\Delta_{\rm S}^{2}-E^{2}}/E$ when
$|E|<\Delta_{\rm S}$.\cite{sun1,sun3,melin,Yeyati} ${\bf \Gamma}^{\rm L/R}=i[{\bf\Sigma}^{r}_{\rm L/R}-{\bf\Sigma}^{a}_{\rm L/R}]$ is the line-width function coupling the left/right semi-infinite terminal with the central scattering region. And the line-width constant of the superconductor terminal is set to be $g_{\rm s}=2\Delta_{\rm S}$ for simplicity.

In the calculation of the spectral function in Fig.~\ref{fig2}\textbf{c}, we consider an infinite long QAHI ribbon covered by the superconductor. Then the momentum $k_x$ is a good quantum number, and ${\bf G}^{r}(E,k_x)=(E{\bf I}-{\bf H}_{k_x}-{\bf\Sigma}^r_{\rm S})^{-1}$.

\textbf{Author Contributions}

Y.Z. and Z.Q. conceived the project; Y.Z. and X.D. prepared the figures; all authors analyzed the main results and wrote the manuscript.

\textbf{Acknowledgements}

We are grateful to Dr. Jie Liu for useful discussion and to Prof. N. E. Davison for polishing the manuscript. Y.Z. is financially supported by NNSFC (11004046 and 11474084), and NSF of Hebei Province of China (A2012205071). Q.S. acknowledges the financial support from NNSFC (11274364) and NBRPC (2012CB921303). Z.Q. is financially supported by the 100 Talents Program of Chinese Academy of Sciences and NNSFC (11474265). The supercomputing center of USTC is acknowledged for computing assistance.

\textbf{Competing financial interests}: The authors declare no competing financial interests.

\clearpage

\end{document}